# Problem Formulation and Fairness

Samir Passi
Information Science
Cornell University
sp966@cornell.edu

Solon Barocas
Information Science
Cornell University
sbarocas@cornell.edu

## ABSTRACT

Formulating data science problems is an uncertain and difficult process. It requires various forms of discretionary work to translate high-level objectives or strategic goals into tractable problems, necessitating, among other things, the identification of appropriate target variables and proxies. While these choices are rarely self-evident, normative assessments of data science projects often take them for granted, even though different translations can raise profoundly different ethical concerns. Whether we consider a data science project fair often has as much to do with the formulation of the problem as any property of the resulting model. Building on six months of ethnographic fieldwork with a corporate data science team—and channeling ideas from sociology and history of science, critical data studies, and early writing on knowledge discovery in databases—we describe the complex set of actors and activities involved in problem formulation. Our research demonstrates that the specification and operationalization of the problem are always negotiated and elastic, and rarely worked out with explicit normative considerations in mind. In so doing, we show that careful accounts of everyday data science work can help us better understand ***how*** and ***why*** data science problems are posed in certain ways—and why specific formulations prevail in practice, even in the face of what might seem like normatively preferable alternatives. We conclude by discussing the implications of our findings, arguing that effective normative interventions will require attending to the practical work of problem formulation.

## CCS CONCEPTS

• **Information systems** → **Data mining;** • **Computing methodologies** → **Machine learning;** • **Human-centered computing** → **Ethnographic studies**

## KEYWORDS

Data Science; Machine Learning; Problem Formulation; Fairness; Target Variable

## 1 INTRODUCTION

Undertaking a data science[1] project involves a series of difficult translations. As Provost and Fawcett point out, "[b]usiness problems rarely are classification problems, or regression problems or clustering problems" [38:293]. They must be ***made*** into questions that data science can answer. Practitioners are frequently charged with turning amorphous goals into well-specified problems—that is, problems faithful to the original business objectives, but also problems that can be addressed by predicting the value of a variable. Often, the outcome or quality that practitioners want to predict—the 'target variable'—is not one that has been well observed or measured in the past. In such cases, practitioners look to other variables that can act as suitable stand-ins—'proxies'.

This process is challenging and far from linear. As Hand argues, "establishing the mapping from the client's domain to a statistical question is one of the most difficult parts of statistical analysis," [22:317] and data scientists frequently devise ways of providing answers to problems that differ from those that seemed to motivate the analysis. In most normative assessments of data science, this work of translation drops out entirely, treating the critical task as one of interrogating properties of the resulting model. However, ethical concerns can extend to the formulation of the problem that a model aims to address, not merely to whether the model exhibits discriminatory effects.

To aid in hiring decisions, for example, machine learning needs to predict a specific outcome or quality of interest. One might want to use machine learning to find "good" employees to hire, but the meaning of "good" is not self-evident. Machine learning requires specific and explicit definitions, demanding that those definitions refer to something measurable. While an employer might want to find personable applicants to join its sales staff, such a quality can be difficult to specify or measure. What counts as personable? And how would employers measure it? Given the challenge of answering these questions, employers might favor a definition focused on sales figures, which they may find easier to monitor. In other words, they might define a "good" employee as the person with the highest predicted sales figures. In so doing, the problem of hiring is formulated as one of predicting applicants' sales figures, not simply identifying "good" employees.

---

[1] The term 'data science,' as used in this paper, refers to the practice of assembling, organizing, processing, modeling, and analyzing data using computational and statistical techniques. In this paper, we focus specifically on supervised machine learning, but the analysis applies more broadly.

As Barocas and Selbst [3] demonstrate, choosing among competing target variables can affect whether a model used in hiring decisions ultimately exhibits a disparate impact. There are three reasons why this might happen. First, the target variable might be correlated with protected characteristics. In other words, an employer might focus on a quality that is distributed unevenly across the population. This alone would not constitute illegal discrimination, as the quality upon which the employer hinges its hiring decisions could be rational and defensible. But, the employer could just as well choose a target variable that is a purposeful proxy for race, gender, or other protected characteristics. This would amount to a form of disparate treatment, but one that might be difficult to establish if the decision rests on a seemingly neutral target variable. The employer could also choose a target variable that seems to serve its rational business interests but happens to generate an avoidable disparate impact—for instance, the employer could choose a different target variable that serves its business objective at least as well as the original choice while also reducing the disparate impact.

Second, the chosen target variable might be measured less accurately for certain groups. For example, arrests are often used as a proxy for crime in applications of machine learning to policing and criminal justice, even though arrests are a racially biased representation of the true incidence of crime [28]. In treating arrests as a reliable proxy for crime, the model learns to replicate the biased labels in its predictions. This is a particularly pernicious problem because the labeled examples in the training data serve as ground truth for the model. Specifically, the model will learn to assign labels to cases similar to those that received the label in the training data, whether or not the labels in the training data are accurate. Worse, evaluations of the model will likely rely on test data that were labeled using the same process, resulting in misleading reports about the model's real-world performance: these metrics would reflect the model's ability to predict the label, not the true outcome. Indeed, when the training and test data have been mislabeled in the same way, there is simply no way to know when the model is making mistakes. Choosing a target variable is therefore often a choice between outcomes of interest that are labeled more or less accurately. When these outcomes are systematically mismeasured by race, gender, or some other protected characteristic, a model designed to predict them will invariably exhibit a discriminatory bias that does not show up in performance metrics.

Finally, different target variables might be more difficult to predict than others depending on the available training data and features. If the ability to predict the target variable varies by population, then the model might subject certain groups to greater errors than others.

Across all three cases, we find that whether a model ultimately violates a specific notion of fairness is often contingent on what the model is designed to predict. Which suggests that we should be paying far greater attention to the choice of the target variable, both because it can be a source of unfairness and a mechanism to avoid unfairness.

**The non-obvious origins of obvious problems.** This might not be surprising because some problem formulations may strike us as obviously unfair. Consider the case of 'financial-aid leveraging' in college admissions—the process by which universities calculate the best possible return for financial aid packages: the brightest students for the least amount of financial aid. To achieve this bargain, the university must predict how much each student is willing to pay to attend the university and how much of a discount would sway an applicant from competitors. In economic terms, 'financial-aid leveraging' calculates each applicant's responsiveness to price, which enables the university to make tailored offers that maximize the likely impact of financial aid on individual enrollment decisions. As Quirk [39] explains: "Take a $20,000 scholarship—the full tuition for a needy student at some schools. Break it into four scholarships each for wealthier students who would probably go elsewhere without the discount but will pay the outstanding tuition if they can be lured to your school. Over four years the school will reap an extra $240,000, which can be used to buy more rich students— or gifted students who will improve the school's profile and thus its desirability and revenue." Such strategies are in effect in schools throughout the United States, and the impact has been an increase in support for wealthier applicants at the expense of their equally qualified, but poorer peers [42,43].

One might, therefore, conclude, as Danielson [11:44] does, that "data mining technology increasingly structures recruiting to many U.S. colleges and universities," and that the technology poses a threat to such important values as equality and meritocracy. Alternatively, one could find, like Cook [9] in a similar thought experiment, that "[t]he results would have been different if the goal were to find the most diverse student population that achieved a certain graduation rate after five years. In this case, the process was flawed fundamentally and ethically from the beginning." For Cook, agency and ethics are front-loaded: a poorly formed question returns undesirable, if correct, answers. Data science might be the enabling device, but the ethical issue precedes the analysis and implementation. The objective was suspect from the start. For Danielson, however, certain ethics seem to flow from data mining itself. Data science is not merely the enabling device, but the impetus for posing certain questions. Its introduction affords new, and perhaps objectionable, ways of devising admissions strategies.

Though they are quite different, these positions are not necessarily incompatible: data science might invite certain kinds of questions, and 'financial-aid leveraging' could be one such example. One might say that data science promotes the formulation of questions that would be better left unasked.

But, this is a strangely unhelpful synthesis: while according agency to the person or people who might formulate the problem, it simultaneously imparts overwhelming influence to the affordances of data science. The effort of getting the question to work as a data science problem drops out entirely, even though this process is where the question actually and ultimately takes shape. The issues of genuine concern—how universities arrive at a workable notion of student quality, how they decide on optimizing for competing variables (student quality, financial burden, diversity, etc.), how the results are put to work in one of many possible ways—are left largely out of view. The indeterminacy of the process, where many of the ethical issues are actually resolved, disappears.

**Problem formulation in practice.** While a focus on the work of problem formulation in real-world applied settings has the potential to make visible the plethora of actors and activities involved in data science work, it has not been the focus of much empirical inquiry to date. We still know very little about the everyday practice of problem formulation. In this paper, we attempt to fill this gap. How and why are specific questions posed? What challenges arise and how are they resolved in everyday practice? How do actors' choices and decisions shape data science problem formulations? Answers to these questions, we argue, can help us to better understand data science as a practice, but also the origin of the qualities of a data science project that raise normative concerns. As researchers work to unpack the normative values at stake in the uses of data science, we offer an ethnographic account of a special financing project for auto lending to make visible the work of problem formulation in applied contexts. In so doing, we show how to trace the ethical implications of these systems back to the everyday challenges and routine negotiations of data science.

In the following sections, we first situate the paper in a longer history attending to the practical dimensions of data science, specifically the task of problem formulation. We then describe our research site and methodology, before moving to the empirical case-study. We conclude by discussing the implications of our findings, positioning the practical work of problem formulation as an important site for normative investigation and intervention.

## 2 BACKGROUND

Our understanding of the role of problem formulation in data science work draws from a long line of research within the history and sociology of science that describes how scientific methods are not just tools for answering questions, but in fact influence the kind of questions we ask and the ways in which we define and measure phenomena [8,25,27,37]. Through different methods, scientists "mathematize" [29,30] the world in specific ways, producing representations that are both contingent *(i.e., they change with a change in methods)* and real *(i.e., they provide actionable ways to analyze the world).* Our practical understanding of a given phenomenon is contingent on the data we choose to represent and measure it with.

The emerging field of critical data studies has brought similar insights to data science: data scientists do not just apply algorithms to data, but work ***with*** algorithms and data—iteratively and often painstakingly—aligning the two together in meaningful ways. Data science work, Passi and Jackson argue [35:2439], "is not merely a collection of formal and mechanical rules, but a situated and discretionary process requiring data analysts to continuously straddle the competing demands of formal abstraction and empirical contingency." Algorithmic results embody specific forms of "data vision"—rule-***based,*** as opposed to rule-***bound,*** applications of algorithms, necessitating judgment-driven work "to apply and improvise around established methods and tools in the wake of empirical diversity" [35:2436].

Data science requires "thoughtful measurement, [...] careful research design, [...and] creative deployment of statistical techniques" [21:80] to identify units of measurement, clean and process data, construct working models, and interpret quantified results [4,5,19,32,34–36]. Subjective decision making is necessary throughout the process. Each of these ***practical*** choices can have profound ***ethical*** implications [10,13,24,31,40], of which data scientists are sometimes well aware. Their everyday work is shot through with "careful thinking and critical reflection" [1:23]. Neff et al. [33], through ethnographic work on academic data science research, show how data scientists often "acknowledge their interpretive contributions" and "use data to surface and negotiate social values." Data, the authors argue, are the starting, and not the end, points in data science.

In academic and research settings—the contexts that inform most of our current understanding of data science—the work of data science comes across mainly as the work of data scientists. Data science projects in applied corporate settings, however, are inherently collaborative endeavors—a world as much of discretion, collaboration, and aspiration as of data, numbers, and models. In such projects, several actors work together to not only make sense of data and algorithmic results but also to negotiate and resolve practical problems. Passi and Jackson [36:18–19], through an ethnography of a corporate data science team, describe how specific issues with data, intuition, metrics, and models pose challenges for corporate data science work, and how organizational actors collaborate, through specific strategies, to manage these problems "in the service of imperfect but ultimately pragmatic and workable forms of analysis." As the authors conclude, "[p]roject managers, product designers, and business analysts are as much a part of applied real-world corporate data science as are data scientists" [36:16].

These strands of research call attention to the role of the work of problem formulation within data science. The relationship between formulated problems and the data we choose to address them is not a one-way street—data are not merely things to answer questions with. Instead, the very formulations of data-driven problems (i.e., the kind of questions we can and do ask) are determined by contingent aspects such as what data are available, what data we consider relevant to a phenomenon, and what method we choose to process them. Problem formulation is as much an outcome of our data and methods as of our goals and objectives. Indeed, defining the data science problem is not only about making the data science process fit specific and specifiable objectives but also making the objectives fit the data science process.

Data miners have long grappled with the role of human judgment and discretion in their practice. The field of Knowledge Discovery in Databases (KDD)—an important predecessor to what we now call data science—emerged to address how choices throughout the data mining process could be formalized in so-called ***process*** models.

## Knowledge Discovery in Databases

**The *iterative* process of applied data mining.** While KDD is commonly associated with data mining and machine learning, the history of the field has less to do with innovations around these techniques than with the process that surrounds their use. Dissatisfied with a lack of applied research in artificial intelligence, scholars and practitioners founded the new sub-field to draw together experts in computer science, statistics, and data management who were interested and proficient in the practical applications of machine learning [see: 18]. The significance of this move owed to a shift in the field's professional focus, not to a change in the substance of its computational techniques. When KDD established itself as an independent field,[2] it also instituted a method for applying machine learning to real-world problems—the KDD process, consisting of a set of computational techniques and specific procedures through which questions are transformed into tractable data mining problems [15,16]. Although the terms KDD and data mining are now used interchangeably—if they are used at all—the original difference between the two is telling. While data mining referred exclusively to the application of machine learning algorithms, KDD referred to the overall process of reworking questions into data-driven problems, collecting and preparing relevant data, subjecting data to analysis, and interpreting and implementing results. The canon of KDD devoted extensive attention not only to the range of problems that lend themselves to machine learning but also to the multi-step process by which these problems can be made into ***practicable*** instances of machine learning. In their seminal paper, Fayyad, Piatetsky-Shapiro, and Smyth, for example, insist on the obvious applicability of data mining while paradoxically attempting to explain and advocate how to apply it in practice—that is, how to make it applicable [14]. KDD covered more than just a set of computational techniques; it amounted to a method for innovating and executing new applications.

The focus on process led to the development of a series of ***process models***—formal attempts to explicate how one progresses through a data mining project, breaking the process into discrete steps [26]. The Cross Industry Standard Process for Data Mining (CRISP-DM) [7], the most widely adopted model, seems to simultaneously ***describe*** and ***prescribe*** the relevant steps in a project's lifecycle. Such an approach grows directly out of the earliest KDD writing. Fayyad, Piatetsky-Shapiro, and Smyth [14] make a point of saying that "data mining is a legitimate activity as long as one understands how to do it," suggesting that there is a particular way to go about mining data to ensure appropriate results. Indeed, the main impetus for developing process models were fears of mistakes, missteps, and misapplications, rather than a simple desire to explicate what it is that data miners do. As Kurgan and Musilek [26] explain, the push "to formally structure [data mining] as a process results from an observation of problems associated with a blind application of [data mining] methods to input data." Notably, CRISP-DM, like the earlier models that preceded it in the academic literature [14–16], emphasized the iterative nature of the process and the need to move back and forth between steps. The attention to feedback loops and the overall dynamism of the process were made especially evident in the widely reproduced visual rendering of the process that adopted a circular form to stress cyclicality [26].

**Negotiated, not faithful, translations.** Business understanding, the first step in the CRISP-DM model, is perhaps the most crucial in a data mining project because it involves the translation of an amorphous problem (a high-level objective or a business goal) into a question amenable to data mining. CRISP-DM describes this step as the process of "understanding the project objectives and requirements from a business perspective [and] then converting this knowledge into a data mining problem definition" [7:10]. This process of 'conversion,' however, is underspecified in the extreme. Translating complex objectives into a data mining problem is not self-evident: "a large portion of the application effort can go into properly formulating the problem (asking the right question) rather than into optimizing the algorithmic details of a particular data-mining method" [14:46]. Indeed, the open-endedness that characterizes such forms of translation work is often described as the 'art' of data mining [12]. Recourse to such terms reveals the degree to which the creativity of the translation process resists its own translation into still more specific parts and processes (i.e., it is artistic only insofar as it resists formalization). But it also highlights the importance of this initial task in determining the very possibility of mining data for some purpose.

CRISP-DM and other practical guidance for data miners or data scientists [see: 6,26] tend to describe problem formulation mainly as part of a project's ***first phase***—an initial occasion for frank conversations between the managers who set strategic business goals, the technologists that manage an organization's data, and the analysts that ultimately work on data. Predictably, those involved in data science work face the difficult challenge of ***faithful translation***—finding the correct mapping between, say, corporate goals, organizational data, and computational problems. Practitioners themselves have long recognized that even when project members reach consensus in formulating the problem, it is a ***negotiated*** translation—contingent on the discretionary judgments of various actors and further impacted by the choice of methods, instruments, and data.

---

[2] Semi-annual workshops started in 1989. Annual conferences began in 1995.

These insights speak to the conditions that motivate data science projects in a way that escapes the kind of technological determinism or data imperative that pervades the current discourses—as if the kinds of questions that data science can answer are always already evident. Getting the automation of machine learning to return the desired results paradoxically involves an enormous amount of manual work and subjective judgment [3]. The work of problem formulation—of iteratively translating between strategic goals and tractable problems—is anything but self-evident, implicated with several practical and organizational aspects of data science work. As Hand points out, "[t]extbook descriptions of data mining tools […] and articles extolling the potential gains to be achieved by applying data mining techniques gloss over [these] difficulties" [23:8].

In the following two sections, we look at the work of data science that is traditionally glossed over. We first describe our research site and methods before moving on to the empirical case-study through which we show how the initial problem formulation comprises a series of *elastic* translations—a set of placeholder articulations that is susceptible to change as the project moves through its many iterations.

## 3 RESEARCH SITE AND METHODS

This paper builds on six months of ethnographic fieldwork with DataVector[3], a multi-billion-dollar US-based e-commerce and new media organization. Established in the 1990s, DataVector owns several companies in domains such as health and automotive. Many of these are multi-million-dollar companies with several thousand clients each. DataVector has a core data science team based on the west coast of the United States that works with companies across different domains. There are multiple teams of data engineers, software developers, and business analysts, both at DataVector and its subsidiaries. One of us worked as a data scientist with the organization's core data science team between June and November 2017, serving as the lead scientist on two corporate data science projects (not reported in this paper) and participating in many others. During ethnographic research, the data science team had eight to eleven members (including one of the authors). The team is headed by Cliff—DataVector's Director of Data Science with 30+ years of industry experience in major technology firms. Cliff and the team report directly to Bill—DataVector's Chief Technology Officer with 20+ years of experience in the technology industry.

During the six-month period, one of us conducted 50+ interviews with data scientists, product managers, business analysts, project managers, and company executives and produced 400+ pages of fieldwork notes and 100+ photographs. Interviews and fieldwork data were transcribed and coded according to the principles of grounded-theory analysis [20,41], inductively analyzing data through several rounds of qualitative analysis. In our analysis, we coded the data in two rounds, focusing on the identification of key categories, themes, and topics as well as the relation between them in the data. While we focus on a specific corporate data science project in this paper, we observed similar dynamics across several other projects. We chose this case because the work of problem formulation was particularly salient in this project.

## 4 CASE-STUDY: SPECIAL FINANCING

CarCorp, a DataVector subsidiary, collects ***special financing*** data: information on people who need car financing but have either low/bad credit scores (between 300-600) or limited credit histories. The company's clientele mainly consists of auto dealers who pay to receive this data (called ***lead data***) that include information such as name, address, mortgage, and employment details (sometimes even the make of the desired automobile). The company collects lead data primarily online: people who need special financing submit their data so that interested dealers can contact them. People requiring special financing face several challenges ranging from the lack of knowledge about available credit sources to difficulties in negotiating interest rates. As liaisons between borrowers and lenders, companies such as CarCorp and its affiliates act as important, sometimes necessary, intermediaries for people requiring special financing. CarCorp serves several dealers across the country.[4] Few dealers collect their own lead data as the money, effort, and technical skills required to do so is enormous. This is a key reason why dealers pay companies such as CarCorp to buy lead data.

CarCorp's technology development and project manager Brian wanted to leverage data science to "improve the quality" of leads. Improving lead quality, Brian argued, will ensure that existing dealers do not churn (i.e., they continue to give their business to CarCorp).

> **Brian (project manager):** "The main goal [is] to improve the quality of our leads for our customers. We want to give actionable leads. […] That is what helps us make money, makes customers continue to use our services" *(Interview, November 1, 2017).*

Initial discussions between the business and data science teams revolved around two themes: (a) defining lead "quality" and (b) finding ways to measure it. Defining lead quality was not straightforward. There were "many stakeholders with different opinions about leads" *(ibid.)*. Some described lead quality as a function of a lead's salary data, while some argued that a lead was good if the dealer had the lead's desired car in their inventory. Everyone on the business team, however, agreed on one thing—as CarCorp's business analyst Ron put it: a "good" lead provided business to the dealer.

> **Ron (business analyst):** "The business team has been talking about [lead quality] for a long time. […] We have narrowed down the lead quality problem to how likely is someone to purchase or to be able to finance a car when

---

[3] Organization, personnel, and project names in this paper have been replaced with pseudonyms to preserve participant anonymity.

[4] The exact number is omitted to preserve company anonymity.

you send them to that dealer?" *(Interview, November 8, 2017).*

Lead "quality" was equated with lead "financeability." It was, however, difficult to ascertain financeability. Different dealers had different special financing approval processes. A lead financeable for one dealer can be, for various reasons, unfinanceable for another. The goal thus was to determine ***dealer-specific financeability*** (i.e., predicting which dealer was most likely to finance a lead). The teams settled on the following definition of "quality": ***a good lead for a dealer was a lead financeable for that dealer.*** This, in turn, framed the problem as one of ***matching leads to dealers that were most likely to finance them.***

CarCorp had a large amount of historical lead data. In 2017 alone, the company had processed close to two million leads. CarCorp, however, had relatively less data on **which** leads had been approved for special financing (let alone data on **why** a lead was approved). The business team asked the data science team to contact data engineers to identify and assess the relevant data sources. The data science team, after investigating the data sources, however, declared that there wasn't enough data on dealer decisions—without adequate data, they argued, it was impossible to match leads with dealers. Few dealers in CarCorp's network shared their approval data with the company. The scarcity of this data stemmed from the challenge of creating business incentives for dealers to provide up-to-date feedback. The incentives for dealers to share information about their approval process with CarCorp were too attenuated.

While the data science team instructed the business team to invest in the collection of up-to-date data on dealer decisions, further discussions ensued between the two teams to find alternate ways to predict dealer-specific financeability using the data that happened to be available already. In debates over the utility of the available data, business analysts and data scientists, however, voiced several concerns ranging from inconsistency (e.g., discrepancies in data values) to unreliability (e.g., distrust of data sources). Business analyst Ron, for instance, was wary of the multiple ways in which the data was collected and generated:

> **Ron (business analyst):** "The entire complex lead ecosystem [...] to outsiders does not make any sense. [...Some data] came from an affiliate. [...] They give a credit score range for a bunch of those leads. So, not exactly the score, but they could say 'this person is between 450-475, and **525**-550,' different buckets. [...] Never realistic, but we pay money, and we have this data. [...] We [also] generate leads organically [online], then there are leads that we [buy from third-parties]. [...] Different pieces of information [are] appended to those leads depending on where it came from" *(Interview, November 8, 2017).*

Only a few CarCorp affiliates augmented lead data with additional information such as credit scores. Dealers run background checks on leads (with their consent) as part of the financing procedure and in the process get a lead's exact credit score from credit bureaus such as Equifax. The Fair Credit Reporting Act (FCRA) prohibits CarCorp from getting a lead's exact credit score from credit bureaus without a lead's explicit consent. Leads have no reason to authorize CarCorp to retrieve their credit data because the company does not make lending decisions; it only collects information about a lead's interest in special financing. CarCorp had to rely on either leads' self-reported credit scores that were collected by a few affiliates or on credit scores provided as part of lead data bought from third-party lending agencies.[5] Business affiliates and third-party agencies provide credit scores in the form of an approximate range (e.g., 476-525). CarCorp had hoped that this data would help them to, for example, differentiate between a subset of leads that appeared identical but exhibited different financeability.

> **Ron (business analyst):** "Two individuals [with] the same age, same income, same housing payment, same everything [...] could have wildly different credit scores. [...] You have those two, and you send them to the same dealer. From our perspective, lead A and B are [...] maybe not exactly [the] same, but close. [...] But, the dealer can finance person A, and they cannot finance person B [...] So, when they [dealers] are evaluating [...] whether they would renew their product with us, if we had sent them a bunch from bucket B, and none from A, they are likely to churn. But, we have no way of knowing who is in bucket A and who is in bucket B. [...] You can have two individuals who measure the same on data points, and they have two different credit scores." *(Interview, November 8, 2017).*

It is not surprising that a lead with a credit score greater than another lead is more likely to secure special financing (even when the leads are otherwise identical). Credit score data is a significant factor in special financing approval processes. CarCorp had this data for about ~10% of their leads (~100,000). While business analysts found it challenging to predict credit score ranges from the other features using traditional statistical analyses, adding credit score ranges in as an additional feature did improve the company's ability to predict lead financeability. The business team wondered if it was possible to use data science to predict credit score ranges for the remaining 90% of the leads (i.e., perhaps machine learning could work where traditional analysis had failed). If successful, credit scores could help ascertain lead financeability—***a financeable lead was a lead with a high credit score.***

The data science team's attempt to assess if they could predict credit scores, however, faced a practical challenge. As mentioned above, credit score data received from affiliates took the form of ranges (e.g., 476-525), not discrete numbers. Different affiliates marked ranges differently. For example, one affiliate may categorize ranges as 476-525, 526-575, etc., while another as 451-500, 501-550, etc. It was not possible to directly use the ranges as class labels for training as the ranges overlapped. The data science team first needed to reconcile different ranges.

As data scientist Alex started working to make the ranges consistent, business analysts came up with a way to make this process easier. Pre-existing market analysis (and, to some extent, word-of-mouth business wisdom) indicated that having a credit score higher than 500 greatly increased a lead's likelihood of obtaining special financing approval. This piece of information had a significant impact on the project. With 500 as the crucial threshold, only two credit score ranges were now significant: *below*-500 and *above*-500. Alex did not need to figure out ways to reconcile ranges such as 376-425 and 401-450 but could bundle them in the *below*-500 category. The *above*-500 credit score range could act as the measure of financeability—*a financeable lead was a lead with a credit score above-500.* The matching problem *(which leads are likely to get financed by a dealer)* was now a classification task *(which leads have a credit score of over 500).* Decreasing the number of classes to two helped attenuate the difficulty of reconciling different ranges but did not help to circumvent it.

---

[5] Third-party lending agencies offer services to help put borrowers in touch with multiple lenders. These agencies get consent from people to perform a soft-pull on their official credit reports from credit bureaus such as Equifax.

> **Alex (data scientist):** If the credit score is below 500, the dealer will simply kill the deal. [...] The problem is there are too many records in the 476-525 range. [...] This makes it difficult *(Fieldwork Notes, June 13, 2017).*

Leads in the 476-525 range were an issue because this range contained not only *below*-500 leads, but also *above*-500 leads. Making mistakes close to the decision boundary is especially consequential for special financing where you want to find people just above the threshold. Alex tried many ways to segregate the leads in this range but the models, according to him, didn't work. Their accuracy, at best, was slightly better than a coin flip. Alex attributed the model's bad performance not only to the presence of leads in the 476-525 range, but also to the limited number of available features (i.e., the data did not present sufficient information for the model to meaningfully differentiate between leads). While the in-house lead dataset was a good starting point, the data scientists knew that accurately classifying leads would require not only creative ways to work with available data, but also a wide variety of data. They had already been scouting for external datasets to augment the in-house lead dataset.

Director of data science Cliff had tasked data science project manager Marcus with the work of finding third-party datasets that could help with classification. Their approach was first to use freely available data to see "how far we get" before buying paid data from companies such as Experian.[6] Free, yet reliable, datasets, however, were hard to come by. Marcus found only a few datasets—Internal Revenue Service (IRS) zip-code-level data on features such as income range, tax returns, and house affordability (i.e., how much an owner might be able to pay for a property). Data scientist Alex tried each dataset but declared that none improved model performance. Members of the data science team wondered if it was worth investing in high-quality paid datasets.

> **Alex:** I used all the data, but the model does not converge.
>
> **Marcus:** What about Experian data? We can get it if you think it will help.

> **Alex:** We will have to buy it first for me to know if it helps with prediction.
>
> **Cliff (director of data science):** Check out the online info on it, and then you and Marcus can figure it out *(Fieldwork Notes, June 16, 2017).*

Without access to the data, Alex argued, it was not possible to clearly know its usefulness. If the data was not going to be helpful, however, it made no sense to buy it—the decision needed to be made based on the available description on Experian's website. They eventually ended up not investing in it. Even after analyzing the dataset's description, Alex was not convinced that the data would increase model performance. Two months later, the project was halted in the absence of actionable progress.

Different actors justified the project's seeming failure in different ways. Data scientist Alex felt that the data was the culprit. Business analyst Ron felt that perhaps the business team unreasonably expected "magic" from data science. For him, the culprit was the nature of the problem itself:

> **Ron (business analyst):** "[It is a] selection-bias problem. We are not dealing with a random sample of the population. [...] These individuals [...] why are they submitting a lead with us? Because it was not easy for them to get financed. By definition, our population is people with at least not great credit, and usually bad credit. Why do you have bad credit? There is like one reason why you have good credit. There are a thousand reasons why you have [...] bad credit. [...] If you show me someone with good credit, I will show you they pay bills on time, have a steady income, etc. If you show me someone with bad credit, they can have a gambling problem, they can be divorced, they could [...] prove income now, but maybe it has been unstable in the past, and we have no way of knowing that. There are literally thousands of reasons that aren't that capture-able" *(Interview, November 8, 2017).*

Business analyst Ron described the failure not in terms of the initially articulated business goal but in terms of the project's current data science problem formulation—not the difficulty of defining the quality of a lead, but the challenge of classifying leads with scores in a specific part of the credit score spectrum. He believed it was possible to classify people with high/low credit scores on the full 300-850 credit score spectrum (e.g., differentiating between a person with a 750 score and a person with a 450 score). He argued, however, that CarCorp's focus on the special financing population meant that the goal was not to classify high/low scores on the full credit spectrum but to demarcate between *different kinds* of low scores on one side of the spectrum (roughly between 300 and 600). Note how Ron, a business analyst, describes the project's failure in relation to a *specific* definition of lead "quality"—it was difficult to know which leads were above or below the credit score threshold of 500. The project was halted when developing an accurate model based on this definition proved impossible. The business and data science team could not figure out any other way to formulate the problem at this stage with the data they had.

---

[6] For example, datasets such as Experian's Premier Aggregated Credit Statistics (http://www.experian.com/consumer-information/premier-aggregated-credit-statistics.html). Even such data only contained aggregated information on credit scores and ranges, and not lead-specific credit scores (which required consent).

## 5 DISCUSSION

Through the above description, we see how the data science problem was formulated differently at different points in the project based on two different sets of targets variables and their possible proxies.

**Proxy #1: Dealer Decisions.** The business team initially described the project goal as the ***improvement of lead quality***—a formulation of what the business team thought the dealers wanted. Note that this goal was in turn related to the broader objective of ***minimizing churn rate***—a formulation of what CarCorp itself wanted. In this way, the problem specification was just as much about keeping clients as it was about satisfying their business objectives. These high-level goals impacted the actors' initial understanding of the project's goal—the quality of leads was seen in relation to dealers and CarCorp's own success. CarCorp decided that if dealers could finance a lead, it was a good lead. The fact that different dealers had different special financing approval processes further impacted the contingent relationship between quality, dealers, and financeability: if a lead was financeable by a specific dealer, it was a good lead for ***that*** dealer. The data science problem, therefore, became the task of ***matching*** leads with dealers that were likely to finance them.

**Proxy #2: Credit Score Ranges.** Data available to support the use of dealer decisions as a proxy, however, were limited. While business analysts did not fully understand how dealers made decisions, they acknowledged, based on market research, the import of credit scores in the special financing approval process—leads with scores higher than 500 were highly likely to get special financing. Credit scores thus became a proxy for a dealer's decision, which was itself a proxy for a lead's financeability, which was, by extension, a proxy for a lead's quality—indeed, a ***chain*** of proxies. The data science problem thus became the task of ***classifying*** leads into below- and above-500 classes.

**Problem formulation is a negotiated translation.** At face value, the relationship between the project's high-level business goal (improving lead quality) and its two different problem formulations (the two sets of target variables and their proxies) may seem like a one-to-many relation—different translations of, in effect, the ***same*** goal. Such an understanding, however, fails to account not only for the amorphous nature of high-level goals (i.e., the difficulty of defining the quality of a lead), but also for the project's iterative and evolving nature (i.e., problem formulations are negotiated, dependent on, for instance, actors' choice of proxy). In our case, actors equated (in order): lead quality with financeability, financeability with dealer decisions, and dealer decisions with credit score ranges. Each of these maneuvers produced different formulations of the objective, in turn impacting actors' articulation and understanding of the project's high-level goal (as seen, for instance, in the way business analyst Ron ultimately accounts for the project's failure).

This is not to argue that the high-level goal to improve lead quality, at some point, transformed into a ***completely different*** objective. Instead, it shows that the translation between high-level goals and tractable data science problems is not a given but a negotiated outcome—stable yet elastic. Throughout the project, the goal of improving lead quality remains recognizably similar but ***practically*** different, evident in the different descriptions, target variables, and proxies for lead quality. Each set of target variables and proxies represents a specific understanding of what a lead's quality is and what it means to improve on it. The quality of a lead is not a preexisting variable waiting to be measured, but an artifact of how our actors define and measure it.

**The values at stake in problem formulation.** Scholars concerned with bias in computer systems have long stressed the need to consider the original objectives or goals that motivate a project, apart from any form of bias that may creep into the system during its development and implementation [17]. On this account, the apparent problem to which data science is a solution determines whether it happens to serve morally defensible ends. Information systems can be no less biased than the objectives they serve.

These goals, however, rarely emerge ready-formed or precisely specified. Instead, navigating the vagaries of the data science process requires reconceiving the problem at hand and making it one that data and algorithms can help solve. In our empirical case, we do not observe a data science project working in the service of an established goal, about which there might be some normative debate. Instead, we find that the normative implications of the project evolve alongside changes in the different problem formulations of lead quality.

On the one hand, for the proxy of dealer-decisions, leads are categorized by their dealer-specific financeability—a lead is only sent to the dealer that is likely to finance them. In formulating the problem as a matching task, the company is essentially ***catering*** to dealer preferences. This approach will recommend leads to dealers that align with the preferences expressed in dealers' previous decisions. In this case, lead financeability operates on a spectrum. Financeability emerges as a more/less attribute: each lead is financeable, some more than others depending on the dealer. Effectively, each lead has at least a chance of being sent to a dealer (i.e., the dealer with the highest probability of financing a lead above some threshold).[7]

On the other hand, for the credit-score proxy, leads are categorized into two classes based on their credit score ranges and only leads with scores greater than 500 are considered financeable. In formulating the problem as the task of classifying leads above or below a score, the company reifies credit scores as the sole marker for financeability. Even if dealers had in the past financed leads with credit scores less than 500, this approach only recommends leads with scores higher than 500, shaping dealers' future financing practices. In this approach, financeability operates as a binary variable: a lead is financeable only if its credit score is higher than 500.

---

[7] Of course, depending on the threshold, some leads will never be sent.

Consequently, leads in the below-500 category may never see the light of day, discounted entirely because the company believes that these leads are not suitable for dealers.

**Different principles; different normative concerns.** Seen this way, the matching version of the problem formulation may appear normatively ***preferable*** to the classification version. But, is this always true? If we prioritize maximizing a person's lending opportunities, the matching formulation of the problem may seem better because it increases a lead's chances of securing special financing. If, however, we prioritize the goal of mitigating existing biases in lending practices (i.e., of alleviating existing dealer biases), the classification problem formulation may come across as the better alternative because it potentially encourages dealers to consider leads different from those they have financed in the past. Through the two scenarios, we see how proxies are not merely ways to equate goals with data but serve to frame the problem in subtly different ways—and raise different ethical concerns as a result.

It is far from obvious which of the two concerns is more serious and thus which choice is normatively preferable—shifting our normative lens alters our perception of fairness concerning the choice of target variables and proxies. In this paper, we have demonstrated how approaching the work of problem formulation as an important site for investigation enables us to have a much more careful discussion about our own normative commitments. This, in turn, provides insights into how we can ensure that projects align with those commitments.

**Always imperfect; always partial.** Translating strategic goals into tractable problems is a labored and challenging process. Such translations do necessary violence to the world that they attempt to model, but also provide actionable and novel ways to address complex problems. Our intention to make visible the elasticity and multiplicity of such translations was thus not to criticize actors' inability to ***find*** the perfectly faithful translation. Quite the opposite: we recognize that translations are always imperfect and partial, and wanted to instead shift the attention to the consequences of different translations and the everyday judgments that drive them.

Our actors, however, did not explicitly debate the ethical implications of their own systems—neither in the way we, as researchers, have come to recognize normative issues, nor in the way we, as authors, have analyzed the implications of their problem formulations in this paper. Practical and organizational aspects such as business requirements, the choice of proxies, the nature of the algorithmic task, and the availability of data impact problem formulations in much more significant and actionable ways than, for instance, the practitioners' normative commitments and beliefs. Indeed, our analysis of the empirical case makes visible how aspects such as analytic uncertainty and financial cost impact problem formulations. For example, the high cost of datasets coupled with the challenge of assessing the data's efficacy without using it made it particularly challenging for actors to leverage additional sources of information.

Yet, as we show in this paper, normative implications of data science systems do ***in fact*** find their roots in problem formulation work—the discretionary judgments and practical work involved in translations between high-level goals and tractable problems. Each translation galvanizes a different set of actors, aspirations, and practices, and, in doing so, creates opportunities and challenges for normative intervention—upstream sites for downstream change. As Barocas et al. [2:6] argue:

> "A robust understanding of the ethical use of data-driven systems needs substantial focus on the possible threats to civil rights that may result from the formulation of the problem. Such threats are insidious, because problem formulation is iterative. Many decisions are made early and quickly, before there is any notion that the effort will lead to a successful system, and only rarely are prior problem-formulation decisions revisited with a critical eye."

If we wish to take seriously the work of unpacking the normative implications of data science systems and of intervening in their development to ensure greater fairness, we need to find ways to identify, address, and accommodate the iterative and less visible work of formulating data science problems—***how*** and ***why*** problems are formulated in specific ways.

## 6 CONCLUSION

In this paper, we focused on the uncertain process by which certain questions come to be posed in real-world applied data science projects. We have shown that some of the most important normative implications of data science systems find their roots in the work of problem formulation. The attempt to make certain goals amenable to data science will always involve subtle transformations of those objectives along the way—transformations that may have profound consequences for the very conception of the problem to which data science has been brought to bear—and what consequently appear to be the most appropriate ways of handling those problems. Thus, the problems we solve with data science are never insulated from the larger process of getting data science to return actionable results. As we have shown, these ends are very much an artifact of a contingent process of arriving at a successful formulation of the problem, and they cannot be easily decoupled from the process at arriving at these ends. In linking the normative concerns that data science has provoked to more nuanced accounts of the on-the-ground process of undertaking a data science project, we have suggested new objects for investigation and intervention: ***which goals are posed and why; how goals are made into tractable questions and working problems;*** and, ***how and why certain problem formulations succeed.***

## ACKNOWLEDGMENTS
The funding for this research was provided by the National Science Foundation grant CHS-1526155, and the Harvard-MIT Ethics and Governance of AI Initiative. We wish to thank our anonymous reviewers for their feedback, and Shira Mitchell, Foster Provost, Malte Ziewitz, and members of Cornell

## REFERENCES

[1] Solon Barocas and Danah Boyd. 2017. Engaging the Ethics of Data Science in Practice. **Commun. ACM** 60, 11 (2017), 23–25.
[2] Solon Barocas, Elizabeth Bradley, Vasant Honavar, and Foster Provost. 2017. Big Data, Data Science, and Civil Rights. *A white paper prepared for the Computing Community Consortium committee of the Computing Research Association*. Retrieved from https://cra.org/ccc/resources/ccc-led-whitepapers/
[3] Solon Barocas and Andrew D. Selbst. 2016. Big Data's Disparate Impact. **104 Calif. Law Rev.** 671, (2016).
[4] danah boyd and Kate Crawford. 2012. Critical Questions for Big Data: Provocations for a cultural, technological, and Scholarly phenomenon. **Information, Commun. Soc.** 15, 5 (June 2012), 662–679.
[5] Lawrence Busch. 2014. A Dozen Ways to Get Lost in Translation: Inherent Challenges in Large Scale Data Sets. **Int. J. Commun.** 8, (2014), 1727–1744.
[6] Kevin Daniel André Carillo. 2017. Let's stop trying to be "sexy" – preparing managers for the (big) data-driven business era. **Bus. Process Manag. J.** 23, 3 (2017), 598–622.
[7] Pete Chapman, Julian Clinton, Randy Kerber, Thomas Khabaza, Colin Shearer, and Rüdiger Wirth. 2000. *CRISP-DM 1.0: Step by step data mining guide*.
[8] Harry M. Collins. 1985. *Changing Order: Replication and Induction in Scientific Practice*. Sage, London.
[9] Jack Cook. 2009. Ethics of Data Mining. In *Encyclopedia of Data Warehousing and Mining*, John Wang (ed.). IGI Global, Hershey, 783–788.
[10] Morgan Currie, Britt S Paris, Irene Pasquetto, and Jennifer Pierre. 2016. The conundrum of police officer-involved homicides: Counter-data in Los Angeles County. **Big Data Soc.** 3, 2 (2016), 1–14.
[11] Peter Danielson. 2009. Metaphors and Models for Data Mining Ethics. In *Social Implications of Data Mining and Information Privacy: Interdisciplinary Frameworks and Solutions*, Ephrem Eyob (ed.). IGI Global, Hershey, 33–47.
[12] Pedro Domingos. 2012. A Few Useful Things to Know About Machine Learning. **Commun. ACM** 55, 10 (2012), 78–87.
[13] Paul Dourish and Edgar Gómez Cruz. 2018. Datafication and data fiction: Narrating data and narrating with data. **Big Data Soc.** 5, 2 (2018), 1–10.
[14] Usama Fayyad, Gregory Piatetsky-Shapiro, and Padhraic Smyth. 1996. From Data Mining to Knowledge Discovery in Databases. **AI Mag.** 17, 3 (1996), 37–54.
[15] Usama Fayyad, Gregory Piatetsky-Shapiro, and Padhraic Smyth. 1996. The KDD Process for Extracting Useful Knowledge from Volumes of Data. **Commun. ACM** 39, 11 (November 1996), 27–34.
[16] William J. Frawley, Gregory Piatetsky-Shapiro, and Christopher J. Matheus. 1992. Knowledge Discovery in Databases: an Overview. **AI Mag.** 13, 3 (1992), 57–70.
[17] Batya Friedman and Helen Nissenbaum. 1996. Bias in computer systems. **ACM Trans. Inf. Syst.** 14, 3 (1996), 330–347.
[18] Mohamed Medhat Gaber (Ed.). 2012. *Journeys to Data Mining: Experiences From 15 Renowned Researchers*. Springer, Berlin, Germany.
[19] Lisa Gitelman. 2006. *Raw Data is an Oxymoron*. MIT Press, MA.
[20] Barney Glaser and Anselm Strauss. 1967. *The Discovery of Grounded Theory: Strategies for Qualitative Research*. Aldine Transactions, Chicago.
[21] Justin Grimmer. 2015. We Are All Social Scientists Now: How Big Data, Machine Learning, and Causal Inference Work Together. **PS Polit. Sci. Polit.** 48, 1 (2015), 80–83.
[22] David Hand. 1994. Deconstructing Statistical Questions. **J. R. Stat. Soc. Ser. A (Statistics Soc.** 157, 3 (1994), 317–356.
[23] David Hand. 2006. Protection or Privacy? Data Mining and Personal Data. In *Advances in Knowledge Discovery and Data Mining*, 1–10.
[24] Lucas D. Introna. 2016. Algorithms, governance, and governmentality: On governing academic writing. **Sci. Technol. Hum. Values** 41, 1 (2016), 17–49.
[25] Bernward Joerges and Terry Shinn. 2001. A Fresh Look at Instrumentation an Introduction. In *Instrumentation Between Science, State and Industry*, Bernward Joerges and Terry Shinn (eds.). Springer Netherlands, Dordrecht, 1–13.
[26] Lukasz A. Kurgan and Petr Musilek. 2006. A survey of Knowledge Discovery and Data Mining process models. **Knowl. Eng. Rev.** 21, 1 (2006), 1–24.
[27] Bruno Latour and Steve Woolgar. 1985. *Laboratory Life: The Construction of Scientific Facts* (2nd ed.). Princeton University Press, Princeton.
[28] Kristian Lum and William Isaac. To predict and serve? **Significance** 13, 5 , 14–19. DOI:https://doi.org/10.1111/j.1740-9713.2016.00960.x
[29] Michael Lynch. 1985. Discipline and the Material Form of Images: An Analysis of Scientific Visibility. **Soc. Stud. Sci.** 15, 1 (1985), 37–66.
[30] Michael Lynch. 1988. The externalized retina: Selection and mathematization in the visual documentation of objects in the life sciences. **Hum. Stud.** 11, (1988), 201–234.
[31] Daniel A. McFarland and H. Richard McFarland. 2015. Big Data and the danger of being precisely inaccurate. **Big Data Soc.** 2, 2 (2015).
[32] Michael Muller, Shion Guha, Eric P.S. Baumer, David Mimno, and N Sadat Shami. 2016. Machine Learning and Grounded Theory Method: Convergence, Divergence, and Combination. In *Proceedings of the 19th International Conference on Supporting Group Work*, 3–8.
[33] Gina Neff, Anissa Tanweer, Brittany Fiore-Gartland, and Laura Osburn. 2017. Critique and Contribute: A Practice-Based Framework for Improving Critical Data Studies and Data Science. **Big Data** 5, 2 (2017), 85–97.
[34] Frank Pasquale. 2015. *The Black Box Society: The Secret Algorithms that Control Money and Information*. Harvard University Press, Cambridge, MA.
[35] Samir Passi and Steven J. Jackson. 2017. Data Vision: Learning to See Through Algorithmic Abstraction. In *Proceedings of the 2017 ACM Conference on Computer Supported Cooperative Work and Social Computing (CSCW '17)*, 2436–2447.
[36] Samir Passi and Steven J. Jackson. 2018. Trust in Data Science: Collaboration, Translation, and Accountability in Corporate Data Science Projects. In *Proceedings of the ACM on Human-Computer Interaction, Vol. 2, CSCW, Article 136*, 1–28.
[37] Trevor J. Pinch and W. E. Bijker. 1984. The Social Construction of Facts and Artefacts: or How the Sociology of Science and the Sociology of Technology might Benefit Each Other. **Soc. Stud. Sci.** 14, 3 (August 1984), 399–441.
[38] Foster Provost and Tom Fawcett. 2013. *Data Science for Business: What You Need to Know About Data Mining and Data-Analytic Thinking*. O'Reilly Media, Sebastopol, CA.
[39] Matthew Quirk. 2005. The Best Class Money Can Buy. **The Atlantic Monthly**.
[40] Gernot Rieder and Judith Simon. 2016. Datatrust: Or, the political quest for numerical evidence and the epistemologies of Big Data. **Big Data Soc.** 3, 1 (2016), 1–6.
[41] Anselm Strauss and Juliet M. Corbin. 1990. *Basics of Qualitative Research: Grounded Theory Techniques and Procedures*. Sage, New York.
[42] Marian Wang. 2013. Public Universities Ramp Up Aid for the Wealthy, Leaving the Poor Behind. **ProPublica**.
[43] Marian Wang. 2014. How Exactly Do Colleges Allocate Their Financial Aid? They Won't Say. **ProPublica**.